\documentstyle[12pt]{article}
\parskip=\medskipamount

\title{Local symmetries in the Hamiltonian framework.\\
      1. Hamiltonian form of the symmetries and the Noether identities.}
\author{A.A. Deriglazov\thanks{alexei@if.ufrj.br ~ On leave of
absence from Dept. of Math. Phys., Tomsk Polytechnical University,
Tomsk, Russia}\\ 
Instituto de F\'\i sica, Universidade Federal do Rio de Janeiro,\\
Rio de Janeiro, Brasil.}
\date{and\\
K.E. Evdokimov\\ 
Department of Physics, Tomsk Polytechnical University,\\ Tomsk, Russia.}
\begin{document}
\maketitle
\large
\begin{abstract}
We study in the Hamiltonian framework the local transformations
$\delta_\epsilon q^A(\tau)=\sum^{[k]}_{k=0}\partial^k_\tau\epsilon^a{}
R_{(k)a}{}^A(q^B, \dot q^C)$ which leave invariant the Lagrangian action:
$\delta_\epsilon S=div$. Manifest form of the symmetry and the
corresponding Noether identities is obtained in the first order
formalism as well as in the Hamiltonian one. The identities has very
simple form and interpretation in the Hamiltonian framework. Part of
them allows one to express the symmetry generators which correspond
to the primarily expressible velocities through the remaining one. 
Other part of the identities allows one to select subsystem of 
constraints with a special structure from the complete constraint 
system. It means,
in particular, that the above written symmetry implies an appearance
of the Hamiltonian constraints up to at least $([k]+1)$ stage. It is
proved also that the Hamiltonian symmetries can always be presented
in the form of canonical transformation for the phase space variables.
Manifest form of the resulting generating function is obtained.
\end{abstract}

\noindent
{\bf PAC codes:}  03.65, 11.15-q\\
{\bf Keywords:} Local symmetries, Hamiltonian systems with constraints.

\section{Introduction}

Formulation of the modern quantum field theory models involves necessarily
an additional nonphysical variables. Their appearance is mostly due to
our desire to incorporate the manifest Poincare invariance and locality  
as the leading principles of the formulation. 
To achieve this, the well known and the standard method
is an appropriate extension of the
physical variable space by the additional degrees of freedom, whose
role is to supply the desired properties of a theory [1-5]. 
In simple cases,
one or more variables of a transparent geometrical origin are needed
(for example, for the case of massive relativistic
particle it is sufficient to introduce only two variables).
In contrast, the modern theories 
incorporate a lot of additional variables, whose nondynamical origin
are supplied either by local symmetries presented in the Lagrangian
action, or by algebraic character of equations of motion for these
variables. In the Hamiltonian framework it manifest himself in
appearance of the constraint system with higher nontrivial 
algebraic structure.

Presence of the additional variables leads to rather
complicated problems on the classical as well as on the quantum level,
and the Hamiltonian methods [1-12] turn out to be well adapted for
investigation of a theory.
Hamiltonization of a Lagrangian system can be formulated as procedure
of rewrittening of the initial dynamics in an equivalent form in terms
of extended (i.e. containing the Lagrangian multipliers) phase space [1].
The result of the procedure (besides the dynamical equations of
motion in the Hamiltonian form) is some system of algebraic
equations, which determines, in particular, physical sector of the theory.
Advantage of the Hamiltonian methods is, in particular, 
that phase space description allows one to separate authomatically
dynamical part of the 
equations of motion from the algebraic one as well as to analyse
arbitrariness of the dynamics for the degenerated theories [4].
It crucially simplifies the problem and is an essential step for the
selfconsistent transition to the quantum theory.

In general case, the above mentioned algebraic system is a mixture of 
the first and the second class constraints as well as of equations for
determining of the Lagrangian multipliers. Also, it can be reducible
(there are exist identities between the equations). The aim of this work
is to study structure of the system in a general framework. We
suppose that there is known local symmetry of the Lagrangian
action, which implies appearance of the corresponding Noether identities
among the equations of motion. Note that it seems to be natural
formulation of
the problem, since the symmetries are usually known for concrete models.
Note also that we do not specify relation among rank of the Hessian
and of the symmetry generators. The theory under consideration can has
first and second class constraints of any stage.
We use the Hamiltonization procedure with the aim to obtain Hamiltonian
form of the symmetries and the corresponding identities. As it will be
shown, the resulting Hamiltonian identities do not involve of the 
time derivatives. In other words, they contain information on the 
algebraic system under consideration. An opposite problem (the problem
of restoring of gauge generators from the known constraint system)
is discussed in [3, 13-22].

For our aims it turns out to be convenient to use the Hamiltonization
procedure in the form developed in [4]. According to this method,
starting from the Lagrangian action one obtains first an equivalent
description for the system in the extended phase space $(q^A, p_A, v^A)$
(first order formalism). Equations of motion which follows from the first
order action contain, in particular, the algebraic one. Part of them
can be solved in the form $v^i=v^i(q,p,v^\alpha)$. Then the Hamiltonian
form of the dynamics [1] can be obtained by means of the direct substitution 
of this solution into all the quantities of the first order formalism.

In this work we repeat these steps for the local Lagrangian symmetry
and for the corresponding Noether identities with the aim to obtain
their form in the Hamiltonian framework. One advantage of this approach
(as compare with discussion based on the Legendre transformation 
[24-26])
is that it turns out to be possible to obtain {\it manifest form} of
the quantities under discussion.

The work is organized as follows. In section 2 we review first steps
of the Hamiltonization procedure [4] with the aim to introduce our
notations. Then we obtain some relations among the Lagrangian
and the Hamiltonian quantities which will be used systematically in
the following sections. In section 3 we illustrate our tricks on example
of the symmetry with at most one derivative acting on parameters (see
also [23] for the case of the symmetry without derivatives).

In the following sections the case of a general local symmetry (see
Eqs.(\ref{77}),(\ref{78}) below) is analysed. In subsections 
4.1, 4.2  we obtain 
manifest form for the corresponding identities in the first order formalism 
as well as in the Hamiltonian one. It is shown that the Noether identities 
in the Hamiltonian form acquires very simple form, their meaning is 
discussed in subsection 4.3. In particular, they allows one to select some 
subsystem of constraints $T$ of the complete constraint system (see 
Eqs.(\ref{99}),(\ref{116})). We prove also that local symmetry with $[k]$ 
derivatives on parameters implies appearance of the Hamiltonian constraints 
up to at least $([k]+1)$ stage. 

In subsections 5.1, 5.2 we obtain manifest form of the local symmetry
in the first order formalism as well as in the Hamiltonian one. The
first order action is invariant under the corresponding transformations
as a consequence of the identities. The same is true for the
Hamiltonian action. We prove also that the Hamiltonian symmetries can
be presented (modulo trivial symmetries of the Hamiltonian action)
in the form of canonical transformation for the phase space variables
(see also [17-22]).
The generating function is find in a manifest form
(see Eq.(\ref{115}) below) and is a
combination of the above mentioned constraints $T$. Results of the
work are enumerated in the Conclusion.

\section{Hamiltonization procedure and some relations among the
Lagrangian and the Hamiltonian quantities.}

Let us consider dynamical system with the action
\begin{eqnarray}\label{1}
S=\int d\tau L(q^A,\dot q^A),
\end{eqnarray}
where $A=1,2,\cdots [A]$. The corresponding equations of motion are
\begin{eqnarray}\label{2}
\frac{\delta S}{\delta q^A}\equiv\frac{\partial L}{\partial q^A}-
\left(\frac{\partial L}{\partial\dot q^A}\right)^{\displaystyle .}=0.
\end{eqnarray}
If will be supposed that the Lagrangian $L$ is at most polynomial on
$\dot q^A$ and is singular
\begin{eqnarray}\label{3}
rank\frac{\partial^2 L}{\partial\dot q^A\partial\dot q^B}\equiv rank 
M_{AB}= [i]<[A]. 
\end{eqnarray}
According to this equation, it is convenient to express the index $A$ as 
$A=(i, \alpha), ~ i=1\cdots [i], ~ \alpha=1\cdots [\alpha]$, where 
$[\alpha]=[A]-[i]$. Without loss of generality [4], 
the matrix $M_{AB}(q, \dot q)$ can be written as follows 
\begin{eqnarray}\label{4}
M_{AB}=\left(\begin{array}{cc}
\ M_{ij} & M_{i\alpha}\\M_{j\beta} & M_{\alpha\beta}
\end{array}\right),
\end{eqnarray}
where $det M_{ij}(q, \dot q)\ne 0$. An opposite matrix will be denoted as 
$\tilde M^{ij}(q, \dot q)$, one has
$M_{ij}\tilde M^{jk}=\delta_i{}^k$.

As the first step of the Hamiltonization procedure, let us rewrite the
theory (1) in terms of the first order action defined on the extended
phase space $(q^A(\tau), ~ p_A(\tau), ~ v^A(\tau))$
\begin{eqnarray}\label{5}
S_v=\int d\tau\left[\bar L(q^A, v^A)+p_A(\dot q^A-v^A)\right],
\end{eqnarray}
where $\bar L=L(q, \dot q)|_{\dot q^A\to v^A}$. 
All the variables are 
considered on equal footing. In particular, one writes equations of motion 
for all of them 
\begin{eqnarray}\label{201}
\dot q^A=v^A\equiv\left\{q^A, \overline{H}(q, p, v^A)\right\}, \cr
\dot p_A=\frac{\partial\bar L}{\partial q^A}\equiv
\left\{p_A, \overline{H}(q, p, v^A)\right\},
\end{eqnarray}
\begin{eqnarray}\label{202}
p_\alpha-\frac{\partial\bar L}{\partial v^\alpha}=0 \quad 
\Longleftrightarrow 
\quad \frac{\partial\overline{H}}{\partial v^\alpha}=0,
\end{eqnarray}
\begin{eqnarray}\label{203}
p_i-\frac{\partial\bar L}{\partial v^i}=0 \quad \Longleftrightarrow
\quad \frac{\partial\overline{H}}{\partial v^i}=0,
\end{eqnarray}
where  $\{ , \}$ is the Poisson bracket and it was denoted
\begin{eqnarray}\label{7}
\overline{H}(q^A, p_A, v^A)\equiv p_Av^A-\bar L(q, v).
\end{eqnarray}
Actions $S$ and $S_v$ describe the same dynamics in the
following sense. \\
a) If $q^A_0(\tau)$ is some solution of the problem (\ref{2}), then the set
of functions $q^A_0$, ~ $v^A_0\equiv\dot q^A_0$, ~
$p_{0A}\equiv\frac{\partial L}{\partial v^A}|_{q_0v_0}$ will be solution
for the system (\ref{201})-(\ref{203}). \\
b) If the set $(q^A_0, p_{0A}, v^A_0)$ is a solution of the system
(\ref{201})-(\ref{203}), then $q^A_0$ obeys to Eq.(\ref{2}).
In accordance with the condition (\ref{3}) one resolves Eq.(\ref{203})
for the multipliers $v^i$ algebraically
\begin{eqnarray}\label{8}
v^i=v^i(q^A, p_j, v^\alpha).
\end{eqnarray}
Hamiltonian form [1] of the initial dynamics can be obtained now by means
of substitution of the Eq.(\ref{8}) into all the quantities of the first
order formalism. We use symbols with bar to denote quantities of the first
order formalism and symbols without bar for the Hamiltonian quantities
\begin{eqnarray}\label{260}
\bar R^A\equiv R^A(q, \dot q)|_{\dot q^A\to v^A}, \cr
R^A\equiv\bar R^A|_{v^i(q^A, p_j, v^\alpha)}.
\end{eqnarray}
 Consider first equations of motion 
(\ref{201})-(\ref{203}). Substitution of Eq.(\ref{8}) into Eq.(\ref{203}) 
gives the identity 
\begin{eqnarray}\label{9}
\Phi_i\equiv p_i-\frac{\partial\bar L}
{\partial v^i}\Biggr|_{v^i}\equiv 0 \quad
\Longleftrightarrow \quad
\frac{\partial\overline{H}}{\partial v^i}\Biggr|_{v^i}\equiv 0,
\end{eqnarray}
while Eq.(\ref{202}) acquire the form
\begin{eqnarray}\label{10}
\Phi_\alpha(q, p)\equiv p_\alpha-\frac{\partial\bar L}{\partial v^\alpha}
\Biggr|_{v^i}=
p_\alpha-f_\alpha(q^A, p_j)=0.
\end{eqnarray}
Note that Eq.(\ref{10}) do not contains the multipliers
$v^\alpha$ (in other case one can use Eq.(\ref{10}) and to express 
some of the multipliers $v^{\alpha^{'}}$ through the remaining one,
in contradiction with the condition (\ref{3})).
Eq.(\ref{10}) determines the primary Hamiltonian constraints
$\Phi_\alpha$.

To find manifest form of equations (\ref{201}) with the multipliers $v^i$
substituted, one introduces Hamiltonian
\begin{eqnarray}\label{11}
H(q^A, p_A, v^\alpha)\equiv\overline{H}|_{v^i}= (p_Av^A-\bar L(q, 
v))|_{v^i}. 
\end{eqnarray}
It obeys the equation
\begin{eqnarray}\label{12}
\frac{dH}{dv^\alpha}=\Phi_\alpha(q, p),
\end{eqnarray}
which can be demonstrated as follows
\begin{eqnarray}\label{13}
\frac{dH}{dv^\alpha}=\frac{\partial\overline{H}}{\partial v^\alpha}
\Biggr|_{v^i}+
\frac{\partial\overline{H}}{\partial v^i}\Biggr|_{v^i}
\frac{\partial v^i}{\partial v^\alpha}=\Phi_\alpha(q, p),
\end{eqnarray}
where Eqs.(\ref{9}),(\ref{10}) were used. General solution of
Eq.(\ref{12}) is
\begin{eqnarray}\label{14}
H(q^A, p_A, v^\alpha)=H_0(q, p)+v^\alpha\Phi_\alpha(q, p),
\end{eqnarray}
where $H_0$ can be find by comparison of equations (\ref{14}) and
(\ref{11})
\begin{eqnarray}\label{15}
H_0(q^A, p_j)=\left(p_iv^i-\bar L(q,v)+ v^\alpha\frac{\partial\bar 
L}{\partial v^\alpha}\right) 
\Biggr|_{v^i(q^A, p_j, v^\alpha)}.
\end{eqnarray}
Nontrivial part of this statement is that $H_0$ do not depends
on the variables $v^\alpha$ (and
$p_\alpha$). Thus we have obtained manifest form of the Hamiltonian
(\ref{14}),(\ref{15}) for an arbitrary Lagrangian action (\ref{1}).
Using the same triks as in Eq.(\ref{13}) one finds also
\begin{eqnarray}\label{16}
\frac{\partial\overline{H}}{\partial q^A}\Biggr|_{v^i}=
\frac{\partial H}{\partial q^A}, \quad
\frac{\partial\overline{H}}{\partial p_A}\Biggr|_{v^i}=
\frac{\partial H}{\partial p_A}.
\end{eqnarray}
It allows one to substitute the functions $v^i(q^A,p_j,v^\alpha)$ into
Eq.(\ref{201}). In the result, Hamiltonian form of the dynamics is
\begin{eqnarray}\label{204}
\dot q^A=\{q^A, H\}, \quad  
\dot p_A=\{p_A, H\},
\end{eqnarray}
\begin{eqnarray}\label{205}
\Phi_\alpha\equiv p_\alpha-f_\alpha(q^A, p_j)=0.
\end{eqnarray}
Note that to obtain Eq.(\ref{204}),(\ref{205}) we have combined in fact
Eq.(\ref{203}) with other
equations of the system (\ref{201})-(\ref{203}). It means that the
systems (\ref{201})-(\ref{203})
and (\ref{204}),(\ref{205}) are equivalent to each other. The first
order action
(\ref{5}) can also be rewritten in the Hamiltonian form. Namely, one notes
that the quantity
\begin{eqnarray}\label{18}
S_H\equiv S_v|_{v^i}=\int d\tau(p_A\dot q^A-H_0-v^\alpha\Phi_\alpha),
\end{eqnarray}
reproduces the equations of motion (\ref{204}), (\ref{205}).

In the following sections we repeat these steps for the local symmetries
and for the
corresponding Lagrangian identities with the aim to obtain manifest form
for these quantities in the Hamiltonian framework.
It is now convenient to enumerate some relations among the Lagrangian and
the Hamiltonian objects which will be used
systematically in the subsequent sections.

Since Eq.(\ref{9}) is identity, one has $\frac{\partial \Phi_i}{\partial 
v^\alpha}\equiv 0$, which can be rewritten as $\frac{\partial v^i}{\partial 
v^\alpha}= -\tilde{\bar M}^{ij}\bar M_{j\alpha}|_{v^i}=-\tilde 
M^{ij}M_{j\alpha}$. On other hand, from equations (\ref{201}) and 
(\ref{204}) it follows $v^i(q^A, p_j, v^\alpha)=\{q^i, H\}$. It can also be 
used for computation of the derivative. Collecting these two results one 
has 
\begin{eqnarray}\label{19}
\frac{\partial v^i}{\partial v^\alpha}=
-\tilde M^{ij}M_{j\alpha}=\{q^i, \Phi_\alpha\}.
\end{eqnarray}
Other derivatives can be obtained in a similar fashion
\begin{eqnarray}\label{20}
\frac{\partial v^i}{\partial p_A}=\tilde M^{ij}\delta^A_j=
\{q^i\{q^j, H\}\}\delta^A_j,
\end{eqnarray}
\begin{eqnarray}\label{21}
\frac{\partial v^i}{\partial q^A}=-\tilde M^{ij}
\frac{\partial^2\bar L}{\partial v^j\partial q^A}\Biggr|_{v^i}=
-\{p_A\{q^i, H\}\}.
\end{eqnarray}
We can use these relations for obtaining derivatives of the constraints
(\ref{10}) as follows
\begin{eqnarray}\label{22}
\frac{\partial\Phi_\alpha}{\partial v^\beta}\equiv 0 ~
\Longrightarrow ~M_{\alpha\beta}-
M_{\alpha i}\tilde M^{ij}M_{j\beta}\equiv 0,
\end{eqnarray}
\begin{eqnarray}\label{23}
\frac{\partial\Phi_\alpha}{\partial p_A}=\delta_\alpha{}^A-
M_{\alpha i}\tilde M^{ij}\delta_j{}^A,
\end{eqnarray}
\begin{eqnarray}\label{24}
\frac{\partial^2\bar L}{\partial q^A\partial v^B}\Biggr|_{v^i}=
-\frac{\partial\Phi_\alpha}{\partial q^A}\delta^\alpha{}_B-
M_{Bi}\frac{\partial v^i}{\partial q^A}.
\end{eqnarray}
Part of equations from (\ref{23}), (\ref{24}) is equivalent to 
Eqs.(\ref{19}), (\ref{21}). Also, for any function $\bar T(q^A, v^A)$ one 
finds 
\begin{eqnarray}\label{25}
\frac{\partial\bar T}{\partial v^i}\Biggr|_{v^i}=
M_{ij}\frac{\partial T}{\partial p_j}\equiv M_{ij}\{q^j, T\}, 
\end{eqnarray}
\begin{eqnarray}\label{26}
\frac{\partial\bar T}{\partial v^\alpha}\Biggr|_{v^i}=
\frac{\partial T}{\partial v^\alpha}+
\frac{\partial\bar T}{\partial v^i}\Biggr|_{v^i}\tilde M^{ij}M_{j\alpha}=
\frac{\partial T}{\partial v^\alpha}+
\frac{\partial T}{\partial p_i}M_{i\alpha},
\end{eqnarray}
and for a set of functions $C_A$
\begin{eqnarray}\label{206}
v^A|_{v^i}C_A=\{q^A, H\}C_A.
\end{eqnarray}
Further, from the condition (\ref{3}) it follows, in particular, that the
matrix $\bar M_{AB}|_{v^i}$ has $[\alpha]$ independent null vectors. Let us
demonstrate that for an arbitrary theory they are
\begin{eqnarray}\label{27}
\frac{\partial\Phi_\alpha}{\partial p_A}\equiv
\left(\delta_\alpha{}^\beta\atop -M_{\alpha j}\tilde M^{ji}\right),
\quad \alpha=1,2,\ldots,[\alpha],
\end{eqnarray}
\begin{eqnarray}\label{28}
\bar M_{BA}|_{v^i}\frac{\partial\Phi_\alpha}{\partial p_A}=0.
\end{eqnarray}
Actually, from Eq.(\ref{23}) one has
\begin{eqnarray}\label{29}
\bar M_{AB}\Biggr|_{v^i}\frac{\partial\Phi_\alpha}{\partial p_B}=
M_{A\alpha}-M_{Ai}\tilde M^{ij}M_{j\alpha}.
\end{eqnarray}
For the case $A=\beta$ the right hand side is zero according to
Eq.(\ref{22}), while for $A=k$ one has $M_{k\alpha}-
M_{ki}\tilde M^{ij}M_{j\alpha}=M_{k\alpha}-M_{k\alpha}=0$, from
which it follows Eq.(\ref{28}). From the right hand side of the 
equality (\ref{27})
it follows that these null vectors are linearly independent.

At last, by using of Eq.(\ref{21}),(\ref{15}),(\ref{16}), some
combinations of the Lagrangian derivatives can be presented
in the Hamiltonian form as follows:
\begin{eqnarray}\label{30}
\frac{\partial\bar L}{\partial q^A}\Biggr|_{v^i}=\{p_A, H\},
\end{eqnarray}
\begin{eqnarray}\label{31}
-\frac{\partial^2\bar L}{\partial q^B\partial v^A}
v^B\bar R^A\Biggr|_{v^i}=
\frac{\partial H}{\partial p_A}\frac{\partial\Phi_\beta}
{\partial q^A}R^\beta+
v^B\frac{\partial v^i}{\partial q^B}M_{iA} R^A. 
\end{eqnarray}
Here $\bar R^A(q, v)$ is any function. In particular, if it is null vector 
of the matrix $\bar M_{AB} : ~ \bar M_{AB}\bar R^B=0$, one has the useful 
relation 
\begin{eqnarray}\label{32}
\bar R^A\left(\frac{\partial\bar L}{\partial q^A}-
\frac{\partial^2\bar L}{\partial q^B\partial v^A}v^B\right)\Biggr|_{v^i}=
R^\alpha\{\Phi_\alpha, H\}.
\end{eqnarray}

\section{Symmetries with at most one derivative acting on parameters.}

Hamiltonization of the local symmetries and the corresponding
Noether identities for
the general case implies sufficiently tedious algebraic manipulations.
So, in this section we demonstrate all the necessary tricks on example
of a symmetry with at most one derivative acting on parameters
(symmetry without derivatives was considered in [23]). Namely,
let us consider infinitesimal local transformations of the form
\begin{eqnarray}\label{40}
\delta_\epsilon q^A=\epsilon^aR_{0a}{}^A(q, \dot q)+
\dot\epsilon^aR_{1a}{}^A(q, \dot q),
\end{eqnarray}
and suppose that the action (\ref{1}) is invariant up to total derivative
term
\begin{eqnarray}\label{41}
\delta_\epsilon S=\int d\tau(\epsilon^a \omega_{0a}+
{\dot\epsilon}^a \omega_{1a})^{\displaystyle .},
\end{eqnarray}
where $\omega_{0a}$, $\omega_{1a}$ are some functions. 
If Eq.(\ref{40}) depends
essentially on all the parameters $\epsilon^a(\tau)$, $a=1,\ldots,[a]$
(namely, if $rank R_{1a}{}^A=[a]$), then $[a]\le[\alpha]$, as it can be
seen from Eq.(\ref{49}) below.

\subsection{Lagrangian identities in the first order formalism.}

Real consequence of the property (\ref{41}) is appearance of identities
among the equations of motion for the theory. To obtain them let us write
Eq.(\ref{41}) in the form of a series on derivatives of $\epsilon^a$
\begin{eqnarray*}
\int d\tau\left[\frac{\partial L}{\partial q^A}R_{0a}{}^A+
\frac{\partial L}{\partial\dot q^A}\dot R_{0a}{}^A\right]\epsilon^a+
\left[\frac{\partial L}{\partial q^A}R_{1a}{}^A+
\frac{\partial L}{\partial\dot q^A}(R_{0a}{}^A+\dot R_{1a}{}^A)\right]
\dot\epsilon^a+ \cr
\ddot\epsilon^a\frac{\partial L}{\partial\dot q^A}R_{1a}{}^A=
\int d\tau(\dot\omega_{0a}\epsilon^a+
(\omega_{0a}+\dot\omega_{1a})\dot\epsilon^a+
\omega_{1a}\ddot\epsilon^a).
\end{eqnarray*}
Since it is fulfiled for an arbitrary $\epsilon^a(\tau)$, one has
\begin{eqnarray}\label{42}
\frac{\partial L}{\partial\dot q^A}R_{1a}{}^A=\omega_{1a},
\end{eqnarray}
\begin{eqnarray}\label{43}
\frac{\partial L}{\partial q^A}R_{1a}{}^A+
\frac{\partial L}{\partial\dot q^A}R_{0a}{}^A+
\frac{\partial L}{\partial\dot q^A}\dot R_{1a}{}^A=
\omega_{0a}+\dot\omega_{1a},
\end{eqnarray}
\begin{eqnarray}\label{44}
\frac{\partial L}{\partial q^A}R_{0a}{}^A+
\frac{\partial L}{\partial\dot q^A}\dot R_{0a}{}^A=\dot\omega_{0a}.
\end{eqnarray}
Substitution of Eq.(\ref{42}) into (\ref{43}) gives expression for
$\omega_{0a}$
\begin{eqnarray}\label{45}
\frac{\delta S}{\delta q^A}R_{1a}{}^A+
\frac{\partial L}{\partial\dot q^A}R_{0a}{}^A=\omega_{0a},
\end{eqnarray}
which can be used in Eq.(\ref{44}) and gives the Noether identities in
the form
\begin{eqnarray}\label{46}
\left(\frac{\delta S}{\delta q^A}R_{1a}{}^A\right)^{\displaystyle .}-
\frac{\delta S}{\delta q^A}R_{0a}{}^A\equiv 0.
\end{eqnarray}
Further, this expression can be presented in the form of a series on
derivatives of $q^A$. It is convenient to introduce the notation
\begin{eqnarray}\label{47}
K_{ia}(q, \dot q)\equiv\left(\frac{\partial L}{\partial q^A}-
\frac{\partial^2 L}{\partial q^B\partial\dot q ^A}\dot q^B\right)R_{ia}{}^A,
\end{eqnarray}
where $i=1,2$. Then the series looks as
\begin{eqnarray}\label{48}
\left[K_{0a}-\dot q^C\frac{\partial}{\partial q^C}K_{1a}\right]- 
\ddot q^A\left[M_{AB}R_{0a}{}^A+\frac{\partial}{\partial\dot q^A}K_{1a}+
\right. \cr
\left.\left(\dot q^C\frac{\partial}{\partial q^C}+
\ddot q^C\frac{\partial}{\partial\dot q^C}\right)M_{AB}R_{1a}{}^B\right]+
\stackrel{(3)}{q}{}^A\left[M_{AB}R_{1a}{}^B\right]\equiv 0.
\end{eqnarray}
Since it is true for any $q^A(\tau)$, the square brackets in Eq.(\ref{48})
must be zero separately. It gives the final form of the Lagrangian identities
for our theory. Since they are fulfiled for any $q^A(\tau)$, they will
remain identities after the substitution
$\dot q^A(\tau)\longrightarrow v^A(\tau)$. {\it In
the result we obtain identities of the first order formalism}
\begin{eqnarray}\label{49}
\bar M_{AB}(q, v)\bar R_{1a}{}^B(q, v)\equiv 0,
\end{eqnarray}
\begin{eqnarray}\label{50}
\bar M_{AB}\bar R_{0a}{}^B+\frac{\partial}{\partial v^B}\bar K_{1a}\equiv 0,
\end{eqnarray}
\begin{eqnarray}\label{51}
\bar K_{0a}-v^B\frac{\partial}{\partial q^B}\bar K_{1a}\equiv 0,
\end{eqnarray}
where $\bar K$ is now function on the extended space
\begin{eqnarray}\label{52}
\bar K_{ia}(q, v)\equiv\left(\frac{\partial\bar L(q, v)}{\partial q^A}-
\frac{\partial^2\bar L}{\partial q^B\partial v^A}v^B\right)
\bar R_{ia}{}^A(q, v).
\end{eqnarray}
Below we demonstrate that Eqs.(\ref{49})-(\ref{51}) supply invariance of
the first order action (\ref{5}) under the corresponding local
transformations (see Eq.(\ref{67})-(\ref{69})).

\subsection{Hamiltonian form of the identities.}

Let us obtain Hamiltonian form of the identities, i.e. we perform
substitution of the multipliers $v^i(q^A,p_j,v^\alpha)$ into
Eqs.(\ref{49})-(\ref{51}).

In acordance with our division of the index: $A=(i, \alpha)$,
Eq.(\ref{49}) can be rewritten as
\begin{eqnarray}\label{53}
\bar R_{1a}{}^i=-\tilde{\bar M}^{ij}\bar M_{j\alpha}\bar R_{1a}{}^\alpha,
\end{eqnarray}
\begin{eqnarray}\label{54}
(\bar M_{\alpha\beta}-\bar M_{\alpha i}\tilde{\bar M}^{ij}\bar 
M_{j\beta})\bar R_{1a}{}^\beta=0, 
\end{eqnarray}
and substitution of the multipliers $v^i$ gives
\begin{eqnarray}\label{55}
R_{1a}{}^i=\{q^i, \Phi_\alpha\}R_{1a}{}^\alpha.
\end{eqnarray}
Eq.(\ref{54}) does not contain new information, see Eq.(\ref{22}).
Similarly, Eq.(\ref{50}) is equivalent to the pair
\begin{eqnarray}\label{56}
R_{0a}{}^i\equiv -\tilde M^{ij}M_{j\alpha}R_{0a}{}^\alpha-
\frac{\partial}{\partial p_i}K_{1a},
\end{eqnarray}
\begin{eqnarray}\label{57}
\frac{\partial}{\partial v^\beta}K_{1a}\equiv 0,
\end{eqnarray}
where Eqs.(\ref{53}),(\ref{54}),(\ref{25}),(\ref{26}) were used.
By using of Eqs.(\ref{19}),(\ref{32}) we find finally
\begin{eqnarray}\label{58}
R_{0a}{}^i\equiv\{q^i, \Phi_\alpha\}R_{0a}{}^\alpha-
\{q^i, R_{1a}{}^\alpha\{\Phi_\alpha, H\}\},
\end{eqnarray}
\begin{eqnarray}\label{59}
\frac{\partial}{\partial v^\beta}\left(R_{1a}{}^{\alpha}
\{\Phi_\alpha, H\}\right)\equiv 0.
\end{eqnarray}
To substitute the multipliers $v^i(q^A, p_j, v^\alpha)$ into the
first term of Eq.(\ref{51}) we use Eqs.(\ref{30}),(\ref{31}),
(\ref{56}), with the result being
\begin{eqnarray}\label{60}
K_{0a}=R_{0a}{}^\alpha\{\Phi_\alpha, H\}+
\frac{\partial H}{\partial q^A}\frac{\partial}{\partial p_A}
(R_{1a}{}^{\alpha}\{\Phi_\alpha, H\})+ \cr
v^B|_{v^i}\frac{\partial v^i}{\partial
q^B}M_{iA}R_{0a}{}^A. 
\end{eqnarray}
For the second term of Eq.(\ref{51}) one has after some algebra
\begin{eqnarray}\label{61}
-\left(v^B\frac{\partial}{\partial q^B}\bar K_{1a}\right)\Biggr|_{v^i}=
-v^B\Biggr|_{v^i}\frac{\partial}{\partial q^B}K_{1a}+
 v^B\frac{\partial v^i}{\partial q^B} 
\left(\frac{\partial}{\partial v^i}\bar K_{1a}\right)\Biggr|_{v^i}= \cr
-\frac{\partial H}{\partial p_A}\frac{\partial}{\partial q^A}
(R_{1a}{}^{\alpha}\{\Phi_\alpha, H\})- v^B|_{v^i}\frac{\partial 
v^i}{\partial q^B}M_{iA}R_{0a}{}^A, 
\end{eqnarray}
where Eqs.(\ref{206}),(\ref{32}),(\ref{50}) were used.
Collecting equations (\ref{60}) and (\ref{61}) one finds finally
the Hamiltonian form of Eq.(\ref{51})
\begin{eqnarray}\label{62}
R_{0a}{}^\alpha\{\Phi_\alpha, H\}-
\left\{R_{1a}{}^\alpha\{\Phi_\alpha, H\}, H\right\}\equiv 0.
\end{eqnarray}
{\it Thus, we have obtained Hamiltonian form of the identities}.
They can be divided on two parts. The
first part means that in arbitrary theory the generators
$R_{1a}{}^i|_{v^i},~R_{0a}{}^i|_{v^i}$ can be expressed through
the remainig one
\begin{eqnarray}\label{64}
R_{1a}{}^i=\{q^i, \Phi_\alpha\}R_{1a}{}^\alpha.
\end{eqnarray}
\begin{eqnarray}\label{63}
R_{0a}{}^i=\{q^i, \Phi_\alpha\}R_{0a}{}^\alpha-
\left\{q^i,~R_{1a}{}^\alpha\{\Phi_\alpha, H\}\right\}.
\end{eqnarray}
The second part involves only the generators $R^\alpha$ and is
\begin{eqnarray}\label{65}
\frac{\partial}{\partial v^\beta}
(R_{1a}{}^{\alpha}\{\Phi_\alpha, H\})\equiv 0,
\end{eqnarray}
\begin{eqnarray}\label{66}
R_{0a}{}^\alpha\{\Phi_\alpha, H\}-
\left\{R_{1a}{}^\alpha\{\Phi_\alpha, H\}, H\right\}\equiv 0.
\end{eqnarray}
Remind that $R_{ia}{}^A\equiv\bar R_{ia}{}^A(q^A, v^A)|_{v^i}$.

\subsection{Local symmetry of the first order action.}

Let us return to discussion of the local symmetries structure.
First we note that {\it the following transformations}
\begin{eqnarray}\label{67}
\delta_\epsilon q^A=\epsilon^a\bar R_{0a}{}^A+\dot\epsilon^a\bar R_{1a}{}^A,
\end{eqnarray}
\begin{eqnarray}\label{68}
\delta_\epsilon p_A=\frac{\partial^2\bar L}{\partial q^A\partial v^B}
\delta_\epsilon q^B+\epsilon^a\frac{\partial}{\partial q^A}\bar K_{1a},
\end{eqnarray}
\begin{eqnarray}\label{69}
\delta_\epsilon v^A=(\delta_\epsilon q^A)^{\displaystyle .},
\end{eqnarray}
{\it leave invariant the first order action (\ref{5}), as a consequence
of the identites (\ref{49})-(\ref{51})}. Actually, variation of the
action $S_v$ under Eqs.(\ref{67}),(\ref{69}) and under some
$\delta p_A$ can be presented as (up to total derivative)
\begin{eqnarray}\label{70}
\delta S_v=\int d\tau ~ \epsilon^a\bar K_{0a}+\dot\epsilon^a\bar K_{1a}-
\dot v^A\bar M_{AB}(\epsilon^a\bar R_{0a}{}^B+
\dot\epsilon^a\bar R_{1a}{}^B)+ \cr
\left(\delta p_A-\frac{\partial^2\bar L}{\partial q^A\partial v^B}
\delta_\epsilon q^B\right)
(\dot q^A-v^A)
\end{eqnarray}
\begin{eqnarray}\label{71}
=\int d\tau ~ \dot\epsilon^av^A\bar M_{AB}\bar R_{1a}{}^B+\epsilon^a
\left(\bar K_{0a}-v^A\frac{\partial}{\partial q^B}\bar K_{1a}\right)- \cr
\epsilon^a\dot v^A\left(\bar M_{AB}\bar R_{0a}{}^B+
\frac{\partial}{\partial v^A}\bar K_{1a}\right)+ \cr
\left(\delta p_A-\frac{\partial^2\bar L}{\partial q^A\partial v^B}
\delta_\epsilon q^B-
\epsilon^a\frac{\partial}{\partial q^A}\bar K_{1a}\right)
(\dot q^A-v^A), 
\end{eqnarray} 
where integration by parts for the second term in Eq.(\ref{70})
was performed. The first and the second lines in Eq.(\ref{71}) are
zero according to Eqs.(\ref{49})-(\ref{51}). Then the variation
$\delta S_v$ will be total derivative if we take $\delta p_A$ according to
Eq.(\ref{68}).

\subsection{Local symmetry of the Hamiltonian action.}

From the discussion in Section 2 one expects that the transformations
(\ref{67})-(\ref{69}) with the multipliers $v^i$ substituted will be
symmetry of the Hamiltonian action (\ref{18}). Let us find their
manifest form. Using Eqs.(\ref{63}),(\ref{64}) one has for the variation
$\delta_\epsilon q^i|_{v^i}$
\begin{eqnarray}\label{72}
\delta_\epsilon q^i|_{v^i}=
\left(\epsilon^aR_{0a}{}^\beta+
\dot\epsilon^aR_{1a}{}^\beta\right)\{q^i, \Phi_\beta\}-
\epsilon^a\left\{q^i, R_{1a}{}^\beta\{\Phi_\alpha, H\}\right\}.
\end{eqnarray}
The variation $\delta_\epsilon q^\alpha|_{v^i}$ can be identically
rewritten in a similar form
\begin{eqnarray}\label{73}
\delta_\epsilon q^\alpha|_{v^i}=
\epsilon^aR_{0a}{}^\alpha+
\dot\epsilon^aR_{1a}{}^\alpha\equiv \cr
\left(\epsilon^aR_{0a}{}^\beta+
\dot\epsilon^aR_{1a}{}^\beta\right)\{q^\alpha, \Phi_\beta\}-
\epsilon^a\left\{q^\alpha, R_{1a}{}^\beta
\{\Phi_\beta, H\}\right\},
\end{eqnarray}
since $\{q^\alpha, \Phi_\beta\}=\delta^\alpha_{{}\beta}$ and since the
quantity $R_{1a}{}^\beta\{\Phi_\beta, H\}$ do not depends on $p_\alpha$. 
For the variation $\delta p_A|_{v^i}$ one has 
\begin{eqnarray}\label{208}
\delta_\epsilon p_A|_{v^i}=
\left(-\frac{\partial\Phi_B}{\partial q^A}\delta_\epsilon q^B+
\epsilon^a\frac{\partial}{\partial q^A}\bar K_{1a}\right)\Biggr|_{v^i}= \cr
-\left(\frac{\partial(\Phi_B|_{v^i})}{\partial q^A}-
\frac{\partial\Phi_B}{\partial v^i}\Biggr|_{v^i}
\frac{\partial v^i}{\partial q^A}\right)\delta_\epsilon q^B|_{v^i}+
\epsilon^a\frac{\partial K_{1a}}{\partial q^A}-
\epsilon^a\frac{\partial v^i}{\partial q^A}
\frac{\partial\bar K_{1a}}{\partial v^i}\Biggr|_{v^i}= \cr
-\frac{\partial\Phi_\alpha}{\partial q^A}\delta_\epsilon q^\alpha
\Biggr|_{v^i}-
M_{Bi}\frac{\partial v^i}{\partial q^A}
\left(\epsilon^aR_{0a}{}^B+
\dot\epsilon^aR_{1a}{}^B\right)+ \cr 
\epsilon^a\frac{\partial}{\partial q^A}
\left(R_{1a}{}^\alpha\{\Phi_\alpha, H\}\right)+ 
\epsilon^aM_{Bi}\frac{\partial v^i}{\partial q^A}
R_{0a}{}^B= \cr
\left(\epsilon^aR_{0a}{}^\alpha+
\dot\epsilon^aR_{1a}{}^\alpha\right)\{p_A, \Phi_\alpha\}-
\epsilon^a\left\{p_A, R_{1a}{}^\alpha\{\Phi_\alpha, H\}\right\}.
\end{eqnarray}
where Eqs.(\ref{9}),(\ref{10}),(\ref{32}),(\ref{49}),(\ref{50}) were
used. {\it Thus we have found Hamiltonian form of the local
symmetry (\ref{40})}
\begin{eqnarray*}
\delta_\epsilon q^A=\{q^A, \Phi_\alpha\}\delta_\epsilon q^\alpha-
\epsilon^a\left\{q^A, R_{1a}{}^\alpha\{\Phi_\alpha, H\}\right\},
\end{eqnarray*}
\begin{eqnarray}\label{74}
\delta_\epsilon p_A=\{p_A, \Phi_\alpha\}\delta_\epsilon q^\alpha-
\epsilon^a\left\{p_A, R_{1a}{}^\alpha\{\Phi_\alpha, H\}\right\},
\end{eqnarray}
\begin{eqnarray*}
\delta_\epsilon v^\alpha=(\delta_\epsilon q^\alpha)^{\displaystyle .},
\end{eqnarray*}
where
\begin{eqnarray}\label{76}
\delta_\epsilon q^\alpha\equiv\epsilon^aR_{0a}{}^\alpha+
\dot\epsilon^aR_{1a}{}^\alpha.
\end{eqnarray}
{\it Hamiltonian action (\ref{18}) is invariant under these transformations,
as a consequence of the identities (\ref{65}),(\ref{66})}. Up to total
derivative, variation of the first term in Eq.(\ref{18}) can be expressed
as follows
\begin{eqnarray*}
\delta_\epsilon(p_A \dot q^A)=\Phi_\alpha(\delta_\epsilon q^\alpha)
^{\displaystyle .}-
\dot\epsilon^aR_{1a}{}^\alpha\{\Phi_\alpha, H\}-
\epsilon^a\dot v^\beta\frac{\partial}{\partial v^\beta}
\left(R_{1a}{}^\alpha\{\Phi_\alpha, H\}\right),
\end{eqnarray*}
while for the second term one has
\begin{eqnarray*}
\delta_\epsilon(-H)=
-\Phi_\alpha(\delta_\epsilon q^\alpha)^{\displaystyle .}+ \cr
\dot\epsilon^aR_{1a}{}^\alpha\{\Phi_\alpha, H\}+
\epsilon^a\left(R_{0a}{}^\alpha\{\Phi_\alpha, H\}-
\{R_{1a}{}^\alpha\{\Phi_\alpha, H\}, \}\right).
\end{eqnarray*}
collecting these terms and using Eqs.(\ref{65}),(\ref{66})
one has $\delta_\epsilon S_H=div$.

\section{Hamiltonization of the identities for the general local symmetry.}

This section is devoted to Hamiltonization of the Lagrangian identities
which correspond to the local symmetry 
\footnote{We show below that presence of the term
$\stackrel{(k)}{\epsilon}$ in
Eq.(\ref{77}) implies appearance of $k$-tiary Hamiltonian constraints.
From this it follows that $[k]<\infty$ for a mechanical system with
finite number of degrees of freedom. Also, with any transformation which
involve variation of the evolution parameter: $\tilde\delta\tau,
\tilde\delta q^A$ one associates unambiguosly the transformations of the
form (\ref{77}) as follows: $\delta\tau=0$,
$\delta q^A=-\dot q^A\tilde\delta\tau+\tilde\delta q^A$. If $\tilde\delta$
is a symmetry of the action, the same will be true for $\delta$. Thus,
Eq.(\ref{77}) incorporates this case also.}
\begin{eqnarray}\label{77}
\delta_\epsilon q^A=\sum^{[k]}_{k=0}{\stackrel{(k)}{\epsilon}}{}^a
R_{(k)a}{}^A(q, \dot q),
\end{eqnarray}
where $\stackrel{(k)}{\epsilon}{}^a\equiv
\frac{d^k}{d\tau^k}\epsilon^a\equiv\partial^k\epsilon^a$.
It is supposed that the action (\ref{1}) is invariant
up to total derivative term
\begin{eqnarray}\label{78}
\delta_\epsilon S=\int d\tau
\left(\sum^{[k]}_{k=0}{\stackrel{(k)}{\epsilon}}{}^a\omega_{ka}\right)
^{\displaystyle .}
\end{eqnarray}
with some functions $\omega_{ka}(q, \dot q)$.

\subsection{\bf{Identites of the first order formalism.}}

The first step is to write Eq.(\ref{78}) in the form of a series on
derivatives of $\epsilon^a$
\begin{eqnarray*}
\int d\tau\left[\frac{\partial L}{\partial q^A}
\sum^{[k]}_{k=0}{\stackrel{(k)}{\epsilon}}{}^aR_{(k)a}{}^A+
\frac{\partial L}{\partial\dot q^A}\sum^{[k]}_{k=0}
\left({\stackrel{(k+1)}{\epsilon}}{}^aR_{(k)a}{}^A+
{\stackrel{(k)}{\epsilon}}{}^a\dot R_{(k)a}{}^A\right)\right]= \cr
\sum^{[k]}_{k=0}\left({\stackrel{(k+1)}{\epsilon}}{}^a\omega_{ka}+
{\stackrel{(k)}{\epsilon}}{}^a\dot\omega_{ka}\right).
\end{eqnarray*}

Since it is fulfiled for any $\epsilon^a(\tau)$, we can compare terms
which are proportional to $\epsilon, \dot\epsilon, \ldots$, separately
\begin{eqnarray}\label{79}
\frac{\partial L}{\partial q^A}R_{([k]+1-k)a}{}^A+
\frac{\partial L}{\partial\dot q^A}\left(R_{([k]-k)a}{}^A+
\dot R_{([k]+1-k)a}{}^A\right)= \cr
\omega_{[k]-k, a}+\dot\omega_{[k]+1-k, a}, \nonumber \\
\frac{\partial L}{\partial q^A}R_{(0)a}{}^A+
\frac{\partial L}{\partial\dot q^A}\dot R_{(0)a}{}^A=\dot\omega_{0a},
\end{eqnarray}
where $k=0,1,\cdots, [k]$, and it is implied
$R_{[k]+1}=\omega_{[k]+1}\equiv 0$.
We can substitute first equation of the system (\ref{79}) into the second
one and so on, it gives manifest form of the functions $\omega_{ka}$. Let
us denote
\begin{eqnarray}\label{80}
S_{(i)a}(q, \dot q, \ddot q)\equiv
\frac{\delta S}{\delta q^A}R_{(i)a}{}^A\equiv  
K_{(i)a}-\ddot q^BM_{BA}R_{(i)a}{}^A, \nonumber \\
K_{(i)a}(q, \dot q)\equiv\left(\frac{\partial L}{\partial q^A}-
\frac{\partial^2L}{\partial q^B\partial\dot q^A}\dot q^B\right)R_{(i)a}{}^A.
\end{eqnarray}
Then one has expressions for $\omega$ as follows
\begin{eqnarray}\label{81}
\sum_{i=0}^{k-1}(-)^{k-1-i}\partial^{k-1-i}S_{([k]-i)a}+
\frac{\partial L}{\partial\dot q^A}R_{([k]-k)a}{}^A=
\omega_{([k]-k)a},
\end{eqnarray}
while the last equation of the system (\ref{79}) gives the Noether
identities
\begin{eqnarray}\label{82}
\sum_{k=0}^{[k]}(-)^{[k]-k}\partial^{[k]-k}S_{([k]-k)a}\equiv 0, \quad
a=1,2,\cdots,[a].
\end{eqnarray}
Note that the $S_{(i)a}$ is at most linear on $\ddot q^A$ and the maximum
possible degree of the time derivative in Eq.(\ref{82}) is $[k]+2$.
Further, Eq.(\ref{82}) can be presented in the form of a series which
consist of the terms \\
$\partial^{k_1} q^{A_1}\cdots\partial^{k_p} q^{A_p}
X_{(A_1\cdots A_p)a}(q, \dot q)$, where the possible values for $p, k_i$ are:
$p=0,1,\cdots,\left[\frac{[k]+2}2\right]$, ~ $k_i\ge 2$,
 ~ $\sum k_i\le[k]+2$, and all the coefficients $X_{(A_1\cdots A_p)a}$ are
symmetric on their indices $A_i$. Since Eq.(\ref{82}) is fulfiled for an
arbitrary $q^A(\tau)$ one concludes
\begin{eqnarray}\label{83}
X_{(A_1\cdots A_p)a}(q, \dot q)\equiv 0.
\end{eqnarray}
Remarkable fact is that only $a\cdot[A]\cdot([k]+1)+a$ functions $X$ from
Eq.(\ref{83}) turns out to be independent. Namely, the direct and tedious
calculation gives the following independent identities which follow form
Eq.(\ref{82})
\begin{eqnarray*}
M_{BA}R_{[k]a}{}^A\equiv 0,
\end{eqnarray*}
\begin{eqnarray}\label{84}
M_{BA}R_{([k]-k)a}{}^A+
\frac{\partial}{\partial\dot q^B}
\left[\sum_{i=1}^k(-)^{i-1}\partial^{i-1}S_{([k]-k+i)a}\right]\equiv 0,
\quad k=1,\ldots,[k],
\end{eqnarray}
\begin{eqnarray*}
\sum_{k=0}^{[k]}(-)^k\dot q^{C_1}\ldots\dot q^{C_k}
\frac{\partial}{\partial q^{C_1}}\ldots\frac{\partial}{\partial q^{C_k}}
K_{(k)a}\equiv 0.
\end{eqnarray*}
This system can be expressed further in an equivalent form in terms of the
quantities $K_{(i)a}$ only. Actually, using the second equation of the
system (\ref{84}) in the third one and so on, one convinces that all the
terms which are proportional to $\ddot q^A$ disappears. This form of the 
identities was obtained also in [26] and used for analysis of constraint 
algebra in a theory without second class constraints. Since the resulting
equations are satisfied for an arbitrary $q^A(\tau)$, they will remain
identites after the substitution $\dot q^A(\tau)\longrightarrow v^A(\tau)$.
In the result, {\it identites of the first order formalism are}
\begin{eqnarray*}
\bar M_{BA}\bar R_{([k])a}{}^A=0,
\end{eqnarray*}
\begin{eqnarray*}
\bar M_{BA}\bar R_{([k]-1)a}{}^A+
\frac{\partial}{\partial v^B}\bar K_{([k])a}=0,
\end{eqnarray*}
\begin{eqnarray*}
\bar M_{BA}\bar R_{([k]-2)a}{}^A+\frac{\partial}{\partial v^B}
\left[\bar K_{([k]-1)a}+(-)^1v^C\frac{\partial}{\partial q^C}\bar K_{([k])a}
\right]=0,
\end{eqnarray*}
\qquad \qquad \qquad \qquad \qquad \qquad \qquad \qquad 
\ldots \\
\begin{eqnarray*}
\bar M_{BA}\bar R_{([k]-k)a}{}^A+ 
\frac{\partial}{\partial v^B}
\left[\sum_{i=1}^{k}(-)^{i-1}v^{C_1}\ldots v^{C_{i-1}}
\frac{\partial}{\partial q^{C_1}}\cdots\frac{\partial}{\partial q^{C_{i-1}}}
\bar K_{([k]-k+i)a}\right]\equiv 0,
\end{eqnarray*}
\qquad \qquad \qquad \qquad \qquad \qquad \qquad \qquad 
\ldots \\
\begin{eqnarray}\label{85}
\bar M_{BA}\bar R_{(0)a}{}^A+\frac{\partial}{\partial v^B}
\left[\sum_{i=1}^{[k]}(-)^{i-1}v^{C_1}\ldots v^{C_{i-1}}
\frac{\partial}{\partial q^{C_1}}\cdots\frac{\partial}{\partial q^{C_{i-1}}}
\bar K_{(i)a}\right]\equiv 0,
\end{eqnarray}
\begin{eqnarray*}
\sum_{k=0}^{[k]}(-)^kv^{C_1}\ldots v^{C_k}
\frac{\partial}{\partial q^{C_1}}\cdots\frac{\partial}{\partial q^{C_k}}
\bar K_{(k)a}\equiv 0,
\end{eqnarray*}
where
\begin{eqnarray}\label{86}
\bar K_{(i)a}\equiv\left(\frac{\partial\bar L}{\partial q^A}-
\frac{\partial^2\bar L}{\partial q^B\partial v^A}v^B\right)
\bar R_{(i)a}{}^A(q, v).
\end{eqnarray}
Similarly to the case which was discussed in section 3, these
identities supply invariance of the first order action under the
corresponding local transformations (see Eq.(\ref{103})-(\ref{105}) below).

Eq.(\ref{85}) promptes the following notation
\begin{eqnarray}\label{87}
\bar T_a{}^{(p)}(q, p, v)\equiv\sum_{i=1}^{p-1}(-)^{i-1}v^{C_1}\ldots
v^{C_{i-1}}\frac{\partial}{\partial q^{C_1}}\ldots
\frac{\partial}{\partial q^{C_{i-1}}}\bar K_{([k]+1-p+i)a} ~.
\end{eqnarray}
Note that the quantities $\bar T_a{}^{(p)}$ can be described as follows: 
let $\bar T_a{}^{(1)}\equiv 0$, then 
\begin{eqnarray}\label{88}
\bar T_a{}^{(p)}=\bar K_{([k]+2-p)a}-
v^B\frac{\partial}{\partial q^B}\bar T_a{}^{(p-1)},
\end{eqnarray}
{\it In this notation our first order identities are}
\begin{eqnarray}\label{89}
\bar M_{BA}\bar R_{([k]+1-p)a}{}^A+
\frac{\partial}{\partial v^B}\bar T_a{}^{(p)}\equiv 0,
\quad p=1,2,\ldots,([k]+1),
\end{eqnarray}
\begin{eqnarray}\label{90}
\bar T_a{}^{([k]+2)}=0.
\end{eqnarray}
Motivation of the "opposite" numeration is as follows: below we show that 
the quantities $T_a{}^{(p)}\equiv\bar T_a{}^{(p)}|_{v^i}$ are some of the 
$p$-ary Hamiltonian constraints. 

\subsection{Hamiltonian form of the identities.}

Since Eq.(\ref{89}),(\ref{90}) are fulfiled for an arbitrary $v^A(\tau)$,
they will remain identities after substitution of the multipliers
$v^i(q^A, p_j, v^\alpha)$ according to Eq.(\ref{8}). This procedure
gives identities of the Hamiltonian formulation. In particular, they
supply invariance of the Hamiltonian action (\ref{18}) under the
corresponding transformations (see Eq.(\ref{108})-(\ref{110}) below)
as well as
contain information on the structure of the Hamiltonian constraint
system. Let us obtain manifest form of the identities. The parts $B=i$
and $B=\alpha$ of Eq.(\ref{89}) are
\begin{eqnarray}\label{91}
\left(\bar M_{ij}\bar R_{([k]+1-p)a}{}^j+
\bar M_{i\beta}\bar R_{([k]+1-p)a}{}^\beta
+\frac{\partial}{\partial v^i}\bar T_a{}^{(p)}\right)\Biggr|_{v^i}=0,
\end{eqnarray}
\begin{eqnarray}\label{92}
\left(\bar M_{\alpha j}\bar R_{([k]+1-p)a}{}^j+
\bar M_{\alpha\beta}\bar R_{([k]+1-p)a}{}^\beta
+\frac{\partial}{\partial v^\alpha}\bar T_a{}^{(p)}\right)\Biggr|_{v^i}=0,
\end{eqnarray}

We can find the generators $R^i$ from Eq.(\ref{91}) and substitute
into Eq.(\ref{92}), the resulting equations are
\begin{eqnarray}\label{93}
R_{([k]+1-p)a}{}^i=-\tilde M^{ij}M_{j\alpha} R_{([k]+1-p)a}{}^\alpha 
-\tilde M^{ij}\left(\frac{\partial}{\partial v^j}\bar T_a{}^{(p)} 
\right)\Biggr|_{v^i}, \cr
M_{\alpha i}\tilde M^{ij}\left(\frac{\partial}{\partial v^j}\bar 
T_a{}^{(p)}\right)\Biggr|_{v^i}- 
\left(\frac{\partial}{\partial v^\alpha}
\bar T_a{}^{(p)}\right)\Biggr|_{v^i}=0,
\end{eqnarray}
where Eq.(\ref{22}) was used. Further, by using of
Eqs.(\ref{19}),(\ref{25}),(\ref{26}) one rewrites Eq.(\ref{93}) as
\begin{eqnarray}\label{94}
R_{([k]+1-p)a}{}^i\equiv
\{q^i, \Phi_\alpha\}R_{([k]+1-p)a}{}^\alpha-
\{q^i, T_a{}^{(p)}\},
\end{eqnarray}
\begin{eqnarray}\label{97}
\frac{\partial}{\partial v^\alpha}T_a{}^{(p)}\equiv 0.
\end{eqnarray}
Note that Eq.(\ref{97}) means, in particular, that $\bar T|_{v^i}$ do not 
depends on $v^\alpha: \\ ~ \bar T_a{}^{(p)}|_{v^i}=T_a{}^{(p)}(q^A, p_j)$. 
Thus we needs to find manifest form for the quantities $\bar 
T_a{}^{(p)}|_{v^i}$.  
Starting from Eq.(\ref{88}) one has 
\begin{eqnarray*}
\bar T_a{}^{(p)}|_{v^i}=\bar K_{([k]+2-p)a}|_{v^i}-
v^B\frac{\partial}{\partial q^B}(\bar T_a^{(p-1)}|_{v^i})- \cr
v^B\frac{\partial v^i}{\partial q^B}
\left(\frac{\partial}{\partial v^i}\bar T_a^{(p-1)}\right)\Biggr|_{v^i}= \cr
\left(\frac{\partial\bar L}{\partial q^A}-
\frac{\partial^2\bar L}{\partial q^B\partial v^A}v^B\right)\Biggr|_{v^i}
R_{([k]+2-p)a}{}^A-
v^B|_{v^i}\frac{\partial}{\partial q^B}T_a^{(p-1)}-  \cr
v^B|_{v^i}\frac{\partial v^i}{\partial q^B}
M_{iA}R_{([k]+2-p)a}{}^A, 
\end{eqnarray*}
as a consequence of Eqs.(\ref{86}),(\ref{89}). The first term can be 
rewritten by means of Eqs.(\ref{30}),(\ref{31}),(\ref{94}) which gives the 
expression (note that the function $\bar T^{(p-1)}|_{v^i}$ do not depends 
on $p_\alpha$) 
\begin{eqnarray*}
R_{([k]+2-p)a}{}^\beta\{\Phi_\beta, H\}+
\frac{\partial H}{\partial q^A}
\frac{\partial}{\partial p_A}T_a^{(p-1)}+
v^B|_{v^i}\frac{\partial v^i}{\partial q^B} M_{iA}R_{([k]+2-p)a}{}^A, 
\end{eqnarray*}
while for the second term one finds
\begin{eqnarray*}
-\frac{\partial H}{\partial p_A}\frac{\partial}{\partial q^A}T_a^{(p-1)}, 
\end{eqnarray*}      
as a consequence of Eq.(\ref{206}). Collecting this terms one
obtain the recurrence relation for determining of the quantities
$T_ a{}^{(p)}$
\begin{eqnarray}\label{98}
T_a{}^{(p)}=R_{([k]+2-p)a}{}^\beta\{\Phi_\beta, H\}+
\{H, T_a^{(p-1)}\},
\end{eqnarray}
Since $T_ a{}^{(2)}$ is known from Eq.(\ref{32}), one finds finally 
\begin{eqnarray}\label{99}
T_a{}^{(p)}(q^A, p_j)=
\sum_{i=2}^p\{ \underbrace{H\{\ldots\{ H}_{\mbox{($i$-2)}} ,
~ R_{([k]+i-p)a}{}^\beta
\{\Phi_\beta, H\}\}\ldots\}\}.
\end{eqnarray}

Thus, {\it Hamiltonian form of the Lagrangian identities (\ref{84}) which
correspond to the general local symmetry (\ref{77}) is}
\begin{eqnarray}\label{100}
R_{([k]+1-p)a}{}^i=\{q^i, \Phi_\alpha\} R_{([k]+1-p)a}{}^\alpha- 
\{q^i, T_a{}^{(p)}\},
\end{eqnarray}
\begin{eqnarray}\label{101}
\frac{\partial}{\partial v^\alpha}T_a^{(p)}=0, \quad
p=2,\ldots,([k]+1),
\end{eqnarray}
\begin{eqnarray}\label{102}
T_a^{([k]+2)}=0,
\end{eqnarray}
with the quantities $T_ a{}^{(p)}$ given by Eq.(\ref{99}). Remind that 
$R_{(k)a}{}^A\equiv \\
\bar R_{(k)a}{}^A(q^A, v^A)|_{v^i}$. Note that the
resulting expressions do not contain of the time derivatives. So, they can 
be compared with the Hamiltonian constraint system.

\subsection{Hamiltonian identities and the complete constraint system.}

The Hamiltonian identites (\ref{101}) states, in particular, that the
quantites $T_a{}^p, ~p=2,3,\ldots,([k]+1)$ are functions of the phase space 
variables $(q^A, p_j)$ only. Their manifest form is 
\begin{eqnarray}\label{116}
T_a^{(2)}=R_{[k]a}{}^\alpha\{\Phi_\alpha, H\}, \cr
\cr
T_a^{(3)}=R_{([k]-1)a}{}^\alpha\{\Phi_\alpha, H\}+
\{H, T_a^{(2)}\}, \cr
\cr
T_a^{(4)}= R_{([k]-2)a}{}^\alpha\{\Phi_\alpha, H\}+\{H, T_a^{(3)}\}, 
\end{eqnarray}
\qquad \qquad \qquad \qquad \qquad \qquad 
\ldots \\
Meaning of these quantites becames clear if we compare Eq.(\ref{116})
with the second, third $\ldots$ stages of the Dirac-Bergmann procedure.
The second stage is to investigate structure of the system
$\{H, \Phi_\alpha\}=0$. It can be rewritten in the canonical form,
one has schematically \footnote{Note that the primary constraints can not
appear on the r.h.s. of this equation, see Eqs.(\ref{10}),(\ref{11}).}
\begin{eqnarray}\label{117}
\{H, \Phi_\alpha\}=Q(q, p)
\left(
\begin{array}{c}
G(q, p, v)\\ \Phi^{(2)}(q, p)\\ 0
\end{array}
\right)=0, \quad
det Q\ne 0,
\end{eqnarray}
where the part $G(q, p, v)=0$ contains (if any) equations for determining 
of the multipliers, while the part $\Phi^{(2)}=0$ determines secondary 
constraints of the theory. From comparison of Eq.(\ref{117}) with the first 
equation of the system (\ref{116}) one concludes that the quantites 
$T_a{}^{(2)}$ are some combinations of the secondary constraints 
$\Phi^{(2)}$. Analogously, the third step is 
\begin{eqnarray}\label{118}
\{H, \Phi^{(2)}\}\Biggr|_{\Phi=\{\Phi, H\}=0}=Q^{'}
\left(
\begin{array}{c}
G^{'}(q, p, v)\\ \Phi^{(3)}(q, p)\\ 0
\end{array}
\right)=0
\end{eqnarray}
which allows one to identity the quantites $T_a{}^{(3)}$ with some of the 
tertiary constraints $\Phi^{(3)}$, and so on. 

{\it Thus, the quantites $T_a{}^{(p)},~p=2,3,\ldots,([k]+1)$ which appears 
in the Hamiltonian identities (\ref{101}) can be identified with some of 
the $p$-ary constraints. Then the identity (\ref{102}) means that 
$([k]+1)$-ary constraints do not create neither $([k]+2)$-ary constraints 
nor equations for determining of the multipliers.} 

From this it follows, in particular, that {\it a theory with the local
symmetry (\ref{77}) necessarily has the Hamiltonian constraints up to
at least $([k]+1)$-stage.}

Note that the subsystem of constraints $T_a{}^{(p)}$ has very special 
structure. Actually, the expression $\{H, T_a{}^{(p)}\}$ gives the 
constraint $T_a{}^{(p+1)}$ modulo the surface $\Phi_\alpha=\{\Phi_\alpha, 
H\}=0$ only. Namely this remarkable structure of the subsystem allows us to 
rewrite the Hamiltonian symmetries in the form of canonical transformation 
(see the subsection 5.2 below). 

\section{Hamiltonian form for the general local symmetry.}

\subsection{The symmetry in the first order formalism.}

In the subsection we prove that {\it the following transformations}
\begin{eqnarray}\label{103}
\delta_\epsilon q^A=\sum_{k=0}^{[k]}\stackrel{(k)}{\epsilon}{}^a
\bar R_{(k)a}{}^A(q, v),
\end{eqnarray}
\begin{eqnarray}\label{104}
\delta_\epsilon p_A=\frac{\partial^2\bar L}{\partial q^A\partial v^B}
\delta_\epsilon q^B+
\sum_{k=0}^{[k]-1}\stackrel{(k)}{\epsilon}{}^a
\frac{\partial}{\partial q^A}\bar T_a^{([k]+1-k)},
\end{eqnarray}
\begin{eqnarray}\label{105}
\delta_\epsilon v^A=(\delta_\epsilon q^A)^{\displaystyle .},
\end{eqnarray}
{\it leave invariant the first order action (\ref{5}) as a consequence of 
the first order identities (\ref{89}),(\ref{90})}. 
Here $\bar T_a{}^{(p)}$ 
is given by Eq.(\ref{87}). Note that the transformations (\ref{77}); 
(\ref{103})-(\ref{105}) and the transformations (\ref{108})-(\ref{110}) 
below are equivalent sets of symmetries in the sense of definition which 
was given in [21]. 

Variation $\delta S_v$ under an arbitrary $\delta q^A, ~ \delta p_A$
and $\delta v^A=(\delta q^A)^{\displaystyle .}$ has the form
\begin{eqnarray}\label{106}
\delta S_v=\int d\tau ~ \left(\frac{\partial\bar L}{\partial q^A}-
\frac{\partial^2\bar L}{\partial q^B\partial v^A}v^B\right)\delta q^A-
\dot v^B\bar M_{BA}\delta q^A+ \cr
\left(\delta p_A-
\frac{\partial^2\bar L}{\partial q^A\partial v^B}\delta q^B\right)
(\dot q^A-v^A).
\end{eqnarray}
After substitution of $\delta q^A$ from Eq.(\ref{103}) into the
first two terms, they can be written as
\begin{eqnarray}\label{107}
\sum_{k=0}^{[k]-1}\stackrel{(k)}{\epsilon}{}^a
(\bar K_{(k)a}-\dot v^B\bar M_{BA}\bar R_{(k)a}{}^A)+
\stackrel{([k])}{\epsilon}{}^a\bar T_a{}^{(2)},
\end{eqnarray}
where the identity with $p=1$ from Eq.(\ref{89}) and Eq.(\ref{88})
were used.
After integration by parts of the last term one has
\begin{eqnarray*}
\sum_{k=0}^{[k]-1}\stackrel{(k)}{\epsilon}{}^a
(\bar K_{(k)a}-\dot v^B\bar M_{BA}\bar R_{(k)a}{}^A)-
\stackrel{([k]-1)}{\epsilon}{}^a
\left(\dot v^B\frac{\partial}{\partial v^B}+
\dot q^B\frac{\partial}{\partial q^B}\right)\bar T_a{}^{(2)}.
\end{eqnarray*}
The intermediate terms form the identity with $p=2$ from
Eq.(\ref{89}), while the remaining one can be presented as
\begin{eqnarray*}
\sum_{k=0}^{[k]-2}\stackrel{(k)}{\epsilon}{}^a
(\bar K_{(k)a}-\dot v^B\bar M_{BA}\bar R_{(k)a}{}^A)+
\stackrel{([k]-1)}{\epsilon}{}^a
\left(\bar K_{([k]-1)a}-v^B\frac{\partial}{\partial q^B}
\bar T_a{}^{(2)}\right)- \cr
\stackrel{([k]-1)}{\epsilon}{}^a
\left(\frac{\partial}{\partial q^B}\bar T_a{}^{(2)}\right)(\dot q^B-v^B),
\end{eqnarray*}
or, equivalently
\begin{eqnarray*}
\sum_{k=0}^{[k]-2}\stackrel{(k)}{\epsilon}{}^a
(\bar K_{(k)a}-\dot v^B\bar M_{BA}\bar R_{(k)a}{}^A)+
\stackrel{([k]-1)}{\epsilon}{}^a\bar T_a^{(3)}- \cr
\stackrel{([k]-1)}{\epsilon}{}^a
\left(\frac{\partial}{\partial q^B}\bar T_a{}^{(2)}\right)(\dot q^B-v^B).
\end{eqnarray*}
Except the last term this expression has the same structure as
Eq.(\ref{107}). Thus, we can repeat these calculations. After $[k]$
integrations by parts and using of the identities (\ref{89}),(\ref{90})
the first two terms in Eq.(\ref{106}) acquire the form
\begin{eqnarray*}
-\sum_{k=0}^{[k]-1}\stackrel{(k)}{\epsilon}{}^a
\left(\frac{\partial}{\partial q^B}\bar T_a^{([k]+1-k)}\right)(\dot q^B-v^B).
\end{eqnarray*}
Thus, modulo of total derivative terms, one has
\begin{eqnarray*}
\delta S_v=\int d\tau ~ \left[
\delta p_A-\frac{\partial^2\bar L}{\partial q^A\partial v^B}
\delta_\epsilon q^B-
\sum_{k=0}^{[k]-1}\stackrel{(k)}{\epsilon}{}^a
\frac{\partial}{\partial q^A}\bar T_a^{([k]+1-k)}\right](\dot q^A-v^A).
\end{eqnarray*}
Then $\delta S_v=div$ if one takes $\delta p_A$ according to Eq.(\ref{104}).

\subsection{The symmetry in the Hamiltonian formalism.}

In this subsection we present manifest form of the transformations
(\ref{103})-(\ref{105}) with the multipliers $v^i$ substituted
according to Eq.(\ref{8}). Then we prove that the resulting
transformations is a symmetry of the Hamiltonian action (\ref{18}).

By using of Eq.(\ref{100}) one finds for the variation
$\delta_\epsilon q^i|_{v^i}$ the expression
\begin{eqnarray*}
\delta_\epsilon q^i|_{v^i}=\sum_{k=0}^{[k]}\stackrel{(k)}{\epsilon}{}^a
\left(R_{(k)a}{}^\beta\{q^i, \Phi_\beta\}-
\{q^i, T_a^{([k]+1-k)}\}\right),
\end{eqnarray*}
where $T_a{}^{(p)}$ is given in Eq.(\ref{99}). The variation
$\delta_\epsilon q^\alpha|_{v^i}$ can be rewritten identically in a similar 
form 
\begin{eqnarray*}
\delta_\epsilon q^\alpha|_{v^i}=
\sum_{k=0}^{[k]}\stackrel{(k)}{\epsilon}{}^aR_{(k)a}{}^\alpha\equiv
\sum_{k=0}^{[k]}\stackrel{(k)}{\epsilon}{}^a
\left(R_{(k)a}{}^\beta\{q^\alpha, \Phi_\beta\}-
\{q^\alpha, T_a^{([k]+1-k)}\}\right),
\end{eqnarray*}
since $\{q^\alpha, \Phi_\beta\}=\delta^\alpha{}_\beta$ and since $T_ 
a{}^{(p)}$
 do not contains of the variable $p_\alpha$. 
Note also that the 
last term in both equations is absent for $k=[k]$, since $T_a{}^{(1)}\equiv 
0$. 

Hamiltonization of the first term in Eq.(\ref{104}) was already
considered, see Eq.(\ref{208})
\begin{eqnarray*}
\frac{\partial^2\bar L}{\partial q^A\partial v^B}\delta q^B\Biggr|_{v^i}=
-\frac{\partial\Phi_\alpha}{\partial q^A}\delta_\epsilon q^\alpha|_{v^i}-
M_{Bi}\frac{\partial v^i}{\partial q^A}\delta_\epsilon q^B\Biggr|_{v^i},
\end{eqnarray*}
while in accordance with Eq.(\ref{89}), the second term of Eq.(\ref{104})
can be presented as
\begin{eqnarray*}
\sum_{k=0}^{[k]-1}\stackrel{(k)}{\epsilon}{}^a
\left(\frac{\partial}{\partial q^A}\bar T_a^{([k]+1-k)}
\right)\Biggr|_{v^i}= \cr
\sum_{k=0}^{[k]-1}\stackrel{(k)}{\epsilon}{}^a
\left[\frac{\partial}{\partial q^A}T_a^{([k]+1-k)}-
\frac{\partial v^i}{\partial q^A}
\left(\frac{\partial}{\partial v^i}
\bar T_a^{[k]+1-k}\right)\Biggr|_{v^i}\right]=\cr
\sum_{k=0}^{[k]-1}\stackrel{(k)}{\epsilon}{}^a
\frac{\partial}{\partial q^A}T_a^{[k]+1-k}+
M_{Bi}\frac{\partial v^i}{\partial q^A}\delta_\epsilon q^B\Biggr|_{v^i}-
\frac{\partial v^i}{\partial q^A}M_{Bi}\stackrel{([k])}{\epsilon}{}^a
R_{[k]a}{}^B.
\end{eqnarray*}
The last term is zero according to Eq.(\ref{89}) with $p=1$. Collecting
these results one has finally
\begin{eqnarray*}
\delta_\epsilon p_A|_{v^i}=
\sum_{k=0}^{[k]}\stackrel{(k)}{\epsilon}{}^a
\left(R_{(k)a}{}^\beta\{p_A, \Phi_\beta\}-
\{p_A, T_a^{([k]+1-k)}\}\right).
\end{eqnarray*}
Thus, {\it the Hamiltonian form for the general local transformations
(\ref{77}) is}
\begin{eqnarray}\label{108}
\delta_\epsilon q^A=\{q^A, \Phi_\alpha\}\delta_\epsilon q^\alpha-
\sum_{k=0}^{[k]-1}\stackrel{(k)}{\epsilon}{}^a
\{q^A, T_a^{([k]+1-k)}\},
\end{eqnarray}
\begin{eqnarray}\label{109}
\delta_\epsilon p_A=\{p_A, \Phi_\alpha\}\delta_\epsilon q^\alpha-
\sum_{k=0}^{[k]-1}\stackrel{(k)}{\epsilon}{}^a
\{p_A, T_a^{([k]+1-k)}\},
\end{eqnarray}
\begin{eqnarray}\label{110}
\delta_\epsilon v^\alpha=(\delta_\epsilon q^\alpha)^{\displaystyle .},
\end{eqnarray}
where
\begin{eqnarray}\label{111}
\delta_\epsilon q^\alpha=\sum_{k=0}^{[k]}\stackrel{(k)}{\epsilon}{}^a
R_{(k)a}{}^\alpha
\end{eqnarray}
and the quantities $T_a{}^{(p)}$ are presented in Eqs.(\ref{98}),
(\ref{99}).

{\it The Hamiltonian identities supply invariance of the Hamiltonian action
(\ref{18}) under these transformations}. Actually, modulo of total
derivative terms, variation of the first term in Eq.(\ref{18}) can be
presented as
\begin{eqnarray*}
\delta_\epsilon(p_A \dot q^A)=
(\delta q^\alpha)^{\displaystyle .}\Phi_\alpha-
\sum_{k=1}^{[k]}\stackrel{(k)}{\epsilon}{}^aT_a^{([k]+2-k)}- \cr
\sum_{k=0}^{[k]-1}\stackrel{(k)}{\epsilon}{}^a
\dot v^\beta\frac{\partial}{\partial v^\beta}T_a^{([k]+1-k)},
\end{eqnarray*}
while for the second term in Eq.(\ref{18}) one finds
\begin{eqnarray*}
\delta_\epsilon(-H)=-(\delta q^\alpha)^{\displaystyle .}\Phi_\alpha+
\sum_{k=0}^{[k]}\stackrel{(k)}{\epsilon}{}^aT_a^{([k]+2-k)}.
\end{eqnarray*}
Collecting these terms one concludes $\delta S_H=div$ as a consequence
of Eqs.(\ref{101}),(\ref{102}).

In conclusion of this subsection let us note that
Eq.(\ref{108})-(\ref{109}) can be presented in the form of canonical
transformations. Let us rewrite them as follows
\begin{eqnarray*}
\delta_\epsilon q^A=\left\{q^A, \delta_\epsilon q^\alpha\Phi_\alpha-
\sum_{k=0}^{[k]-1}\stackrel{(k)}{\epsilon}{}^a
T_a^{([k]+1-k)}\right\}-
\{q^A, \delta_\epsilon q^\alpha\}\Phi_\alpha,
\end{eqnarray*}
\begin{eqnarray}\label{112}
\delta_\epsilon p_A=\left\{p_A, \delta_\epsilon q^\alpha\Phi_\alpha-
\sum_{k=0}^{[k]-1}\stackrel{(k)}{\epsilon}{}^a
T_a^{([k]+1-k)}\right\}-
\{p_A, \delta_\epsilon q^\alpha\}\Phi_\alpha,
\end{eqnarray}
\begin{eqnarray*}
\delta_\epsilon v^\alpha=(\delta_\epsilon q^\alpha)^{\displaystyle .}.
\end{eqnarray*}
It can be acompanied by the trivial transformation of the
Hamiltonian action (see [23])
\begin{eqnarray*}
\bar\delta_\epsilon q^A=\{q^A, \delta_\epsilon q^\alpha\}\Phi_\alpha,
\end{eqnarray*}
\begin{eqnarray}\label{113}
\bar\delta_\epsilon p_A=\{p_A, \delta_\epsilon q^\alpha\}\Phi_\alpha,
\end{eqnarray}
\begin{eqnarray*}
\bar\delta_\epsilon v^\alpha=-(\delta_\epsilon q^\alpha)^{\displaystyle .}+
\{H, \delta_\epsilon q^\alpha\}.
\end{eqnarray*}
Combination of the equations (\ref{112}) and (\ref{113}) has the desired
canonical form.

Thus we have proved that the Hamiltonian symmetries which correspond to a
general local symmetry (\ref{77}) of the Lagragian formalism can be
presented in the form of canonical transformation for the phase space
variables $q^A,~p_A$
\begin{eqnarray}\label{114}
\delta_\epsilon q^A=\{q^A, G\}, \quad
\delta_\epsilon p_A=\{p_A, G\}, \cr
\delta_\epsilon v^\alpha=\{H, \delta_\epsilon q^\alpha\},
\end{eqnarray}
where the generating function is
\begin{eqnarray}\label{115}
G=\sum_{k=0}^{[k]}\stackrel{(k)}{\epsilon}{}^a
R_{(k)a}{}^\alpha\Phi_\alpha-
\sum_{k=0}^{[k]-1}\stackrel{(k)}{\epsilon}{}^a
T_a^{([k]+1-k)}.
\end{eqnarray}

\section{Conclusion}

Let us enumerate results of this work.

1. ~ Starting from the Lagrangian theory with the local symmetry of a
general form
\begin{eqnarray}\label{120}
\delta_\epsilon q^A=\sum^{[k]}_{k=0}{\stackrel{(k)}{\epsilon}}{}^a
R_{(k)a}{}^A(q, \dot q),
\end{eqnarray}
we have obtained {\it manifest form} of the symmetry and of the
corresponding
Noether identities in the first order formalism (Eqs.(\ref{103})-
(\ref{105}),(\ref{89});(\ref{90})) as well as in the Hamiltonian
one (Eqs.(\ref{108})-
(\ref{110}), (\ref{100})-(\ref{102})). The identities supply invariance
of the first order action and of the Hamiltonian one under the
corresponding transformations.

2. ~ The Hamiltonian identities consist of two parts. The first part allows
one to express the generators $R^i$ through the others (remind that now 
$R_{(k)a}{}^A\equiv\bar R_{(k)a}{}^A(q^A, v^A)|_{v^i}$ and $R^A=(R^i, 
R^\alpha)$ 
in accordance with Eq.(\ref{3})) 
\begin{eqnarray}\label{121}
R_{([k]+1-p)a}{}^i=\{q^i, \Phi_\alpha\} R_{([k]+1-p)a}{}^\alpha- 
\{q^i, T_a{}^{(p)}\}.
\end{eqnarray}
Manifest form of the quantities $T$ is known
\begin{eqnarray}\label{122}
T_a{}^{(p)}(q^A, p_j)=
\sum_{i=2}^p\{ \underbrace{ H\{\ldots\{ H}_{\mbox{($i$-2)}} ,
~ R_{([k]+i-p)a}{}^\beta
\{\Phi_\beta, H\}\}\ldots\}\}.
\end{eqnarray}

3. ~ The second part of the Hamiltonian identities
\begin{eqnarray}\label{123}
\frac{\partial}{\partial v^\alpha}T_a^{(p)}\equiv 0, \quad
p=2,\ldots,([k]+1),
\end{eqnarray}
\begin{eqnarray}\label{124}
T_a^{([k]+2)}\equiv 0,
\end{eqnarray}
has the following interpretation. Eq.(\ref{123}) means, in particular, that 
the quantities $T_a{}^{(p)}$ do not depend on $v^\alpha$. We have 
demonstrated also that the quantities $T_a{}^{(p)}(q^A, p_j)$ are some part 
of the $p$-ary Hamiltonian constraints. Then Eq.(\ref{123}) states that in 
a theory with the local symmetry (\ref{120}) the complete constraint system 
contains subsystem of constraints $T_a{}^{(p)}\approx 0$ of the 
special structure (\ref{122}). Equation (\ref{124}) means that 
$([k]+1)$-ary constraints $T$ do not create neither new constraints nor 
equations for determining of the multipliers. 

4. ~ Local symmetry (\ref{120}) for the Lagrangian theory
implies appearance of the Hamiltonian constraints up to at least
$([k]+1)$ stage.

5. ~ We have proved that the Hamiltonian symmetry (\ref{108})-(\ref{110}),
which corresponds to the Lagrangian one (\ref{120}), can be presented
in the form of canonical transformation (for the phase space variables
$q^A, p_A$)
\begin{eqnarray}\label{125}
\delta_\epsilon q^A=\{q^A, G\}, \quad
\delta_\epsilon p_A=\{p_A, G\}, \cr
\delta_\epsilon v^\alpha=\{H, \delta_\epsilon q^\alpha\},
\end{eqnarray}
where the generating function is the following combination of the primary
constraints $\Phi_\alpha$ and the constraints $T_a{}^{(p)}$
\begin{eqnarray}\label{126}
G=\sum_{k=0}^{[k]}\stackrel{(k)}{\epsilon}{}^a
R_{(k)a}{}^\alpha\Phi_\alpha-
\sum_{k=0}^{[k]-1}\stackrel{(k)}{\epsilon}{}^a
T_a^{([k]+1-k)}.
\end{eqnarray}
Difference among the equations (\ref{108})-(\ref{110}) and the canonical
transformations (\ref{125}) is the trivial symmetry of the Hamiltonian action
\begin{eqnarray}\label{127}
\bar\delta_\epsilon q^A=\{q^A, \delta_\epsilon q^\alpha\}\Phi_\alpha,
\quad
\bar\delta_\epsilon p_A=\{p_A, \delta_\epsilon q^\alpha\}\Phi_\alpha,\cr
\bar\delta_\epsilon v^\alpha=-(\delta_\epsilon q^\alpha)^{\displaystyle .}+
\{H, \delta_\epsilon q^\alpha\}.
\end{eqnarray}From the previous discussion one expects that 
the constraints $T_a{}^{(p)}$ 
are of first class in the complete constraint system. Also, from Eqs. 
(\ref{122})-(\ref{124}) it follows that for the simple cases (symmetry 
without derivative or with one derivative only) the gauge generators $R$ 
can be easily restored starting from the Hamiltonian formulation. We hope 
that the results obtained allows one to formulate a simplified procedure 
(as compare with [14, 21]) for the general case also. These problems will 
be discussed in a forthcoming paper.

\section*{Acknowledgments.}

One of the authors (A.A.D.) thanks D.M. Gitman for useful discussions.
The work of A.A.D. has been supported by FAPERJ and partially by
DFG-RFBR project No 99-02-04022, and by Project INTAS-96-0308.


\begin{thebibliography}{nn}
\bibitem{1} P.A.M. Dirac, Can. J. Math. {\bf 2}
(1950) 129; Lectures on Quantum Mechanics (Yeshiva Univ., New York, 1964). 
\bibitem{2} J.L. Anderson and P.G. Bergmann, Phys. Rev. {\bf 83} (1951) 1018.
\bibitem{3} P.G. Bergmann and I. Goldberg,
Phys. Rev. {\bf 98} (1955) 531.
\bibitem{4} D. M. Gitman and I. V. Tyutin, Quantization of Fields
with Constraints (Berlin: Springer-Verlag, 1990).
\bibitem{5} M. Henneaux and C. Teitelboim, Quantization of Gauge Systems
(Princeton: Princeton Univ. Press, 1992).
\bibitem{6} K. Kamimura, Nuovo Cimento {\bf 68B} (1982) 33. 
\bibitem{7} C. Batlle, J. Gomis, J.M. Pons and N. Roman-Roy,
J. Math. Phys. {\bf 27} (1986) 12.
\bibitem{8} A. Cabo and D. Louis-Martinez, Phys. Rev. {\bf D42} (1990) 2726.
\bibitem{9} R. Banerjee and J. Barcelos-Neto, Ann. Phys. {\bf 265} (1998) 134.
\bibitem{10} J. Barcelos-Neto, Phys. Rev. {\bf D55} (1997) 2265.
\bibitem{11} C. Wotzasek, Ann. Phys. {\bf 243} (1995) 76.
\bibitem{12} J.M. Pons and J.Antonio Garc\'\i a, hep-th/9908151. 
\bibitem{13} J.Gomis, K. Kamimura and J.M. Pons, Euruphys. Lett., {\bf 2} 
(1986) 187.
\bibitem{14}  M. Henneaux, C. Teitelboim and J. Zanelli, Nucl. Phys. 
{\bf B332} (1990) 169.
\bibitem{15} V. Mukhanov and A. Wipf, Int. J. Mod. Phys., {\bf A10 } 
(1995) 579.
\bibitem{16} R. Sugano and T. Kimura, Phys. Rev. {\bf D41} (1990) 41.
\bibitem{17} R. Sugano and T. Kimura, J. Math. Phys. {\bf 31} (1990) 2337.
\bibitem{18} J. Gomis, M. Henneaux and J.M. Pons, Class. Quantum. Grav. 
{\bf 7} (1990) 1089.
\bibitem{19} S.A. Gogilidze, V.V. Sanadze, Yu.S. Surovtsev and 
F.G. Tkebuchava, Theor. Math. Phys. {\bf 102} (1995) 47.
\bibitem{20} V.A. Borovkov and I.V. Tyutin, Physics of Atomic Nuclei
{\bf 61} (1998) 1603.
\bibitem{21} V.A. Borovkov and I.V. Tyutin, Physics of Atomic Nuclei
{\bf 62} (1999) 1070.
\bibitem{22} R. Banerjee, H.J. Rothe and K.D. Rothe, hep-th/9907217;
hep-th/9909039.
\bibitem{23} A.A. Deriglazov, hep-th/9412244.
\bibitem{24} Kh.S. Nirov and A.V. Razumov, Int. J. Mod. Phys. {\bf 7} 
(1992) 5549.
\bibitem{25} Kh.S. Nirov and A.V. Razumov, J. Math. Phys. {\bf 34} 
(1993) 3933. 
\bibitem{26} Kh.S. Nirov, Int. J. Mod. Phys. {\bf A 10} (1995) 4087.
\end{thebibliography}
\end{document}